\documentclass[12pt]{article}
\usepackage{amssymb}
\usepackage{amsmath}
\usepackage{amstext}
\usepackage{graphicx,epsfig}
\usepackage{epsfig}
\usepackage{color}
\usepackage{parskip}
\usepackage{cite}

\newcommand{\Comment}[1]{{}}
\definecolor{MyDarkBlue}{rgb}{0.15,0.15,0.45}
\usepackage[linktocpage=true]{hyperref}
\hypersetup{
colorlinks=true,
citecolor=MyDarkBlue,
linkcolor=MyDarkBlue,
urlcolor=MyDarkBlue,
pdfauthor={Kurt Hinterbichler},
pdftitle={},
pdfsubject={hep-th}
}

\newcommand{\be}{\begin{equation}}
\newcommand{\ee}{\end{equation}}
\newcommand{\bal}{\begin{align}}
\newcommand{\eal}{\end{align}}
\newcommand{\bals}{\begin{align*}}
\newcommand{\eals}{\end{align*}}
\newcommand{\bea}{\begin{eqnarray}}
\newcommand{\eea}{\end{eqnarray}}
\newcommand{\beas}{\begin{eqnarray*}}
\newcommand{\eeas}{\end{eqnarray*}}
\newcommand{\nn}{\nonumber}

\def\({\left(}
\def\){\right)}

\numberwithin{equation}{section}

\begin{document}


\begin{center}
{\LARGE {Kaluza-Klein Reduction of Massive and Partially Massless Spin-2 Fields \\ \vspace{.2cm} }}
\end{center}
\vspace{2truecm}
\thispagestyle{empty} \centerline{
{\large {James Bonifacio,}}$^{a,}$\footnote{E-mail: \Comment{\href{mailto:james.bonifacio@physics.ox.ac.uk}}{\tt james.bonifacio@physics.ox.ac.uk}}
{\large { Kurt Hinterbichler$^{b,}$}}\footnote{E-mail: \Comment{\href{mailto:kurt.hinterbichler@case.edu}}{\tt kurt.hinterbichler@case.edu}}
}

\vspace{1cm}

\centerline{{\it
$^{a}$Theoretical Physics, University of Oxford,}}
\centerline{{\it DWB, Keble Road, Oxford, OX1 3NP, UK}}

\vspace{.3cm}

\centerline{\it ${}^{\rm b}$CERCA, Department of Physics, Case Western Reserve University, }
\centerline{\it 10900 Euclid Ave, Cleveland, OH 44106, USA}

\begin{abstract}

We describe the dimensional reduction of massive and partially massless spin-2 fields on general Einstein direct product manifolds.  
As with massless fields, the higher-dimensional gauge symmetry of the partially massless field displays itself upon dimensional reduction as a tower of St\"uckelberg symmetries for the massive modes of the tower.  Unlike the massless case, the zero mode of the gauge symmetry does not display itself as a lower-dimensional non-St\"uckelberg gauge symmetry enforcing partial masslessness on the zero mode.  Partial masslessness is destroyed by the dimensional reduction and the zero mode gauge symmetry instead serves to eliminate the radion.
In addition, we study the fully non-linear dimensional reduction of dRGT massive gravity on a circle, which results in a massive scalar-tensor-vector theory which we expect to be ghost-free, and whose scalar-tensor sector is a special case of mass-varying massive gravity.

\end{abstract}

\newpage
\tableofcontents
\newpage
\renewcommand*{\thefootnote}{\arabic{footnote}}
\setcounter{footnote}{0}

\section{Introduction}
\parskip=5pt
\normalsize

Since Kaluza
and Klein \cite{Kaluza:1921tu,Klein:1926tv}, the idea of compact extra dimensions has played a crucial role in many developments of theoretical physics.  This is true whether one believes the extra dimensions are to be thought of as physically real, as in phenomenological studies of string compactifications or large extra dimensions \cite{ArkaniHamed:1998rs}, or are to be thought of as mathematical tools used to construct or explore purely four-dimensional theories\footnote{For example, in maximally supersymmetric Yang-Mills theory, a weakly coupled Lagrangian description can be most easily constructed by starting with a simpler action in 10 dimensions and reducing on a torus \cite{Brink:1976bc}, and at strong coupling, a 10-dimensional space compactified on a 5-dimensional sphere plays a central role through the AdS/CFT correspondence \cite{Maldacena:1997re}.}.

The hallmark of any Kaluza-Klein (KK) reduction on a compact manifold is an infinite but discrete tower of modes of increasing masses.  These masses are determined by the eigenvalues of appropriate Laplacians acting on the internal space. 

In the case where the higher-dimensional fields are massless fields with spin, there are gauge symmetries.  These gauge symmetries also get dimensionally reduced and become an infinite tower of lower-dimensional gauge symmetries.  Most of these symmetries become algebraic St\"uckelberg-like fields associated with the massive modes in the KK tower.   These are gauged shift symmetries which when fixed make the higher KK modes eat their corresponding longitudinal Goldstone bosons to become massive.  The exception is when there are zero modes.  In this case, there are lower-dimensional massless fields at the bottom of the KK tower, and the zero modes of the gauge symmetry are not St\"uckelberg but instead become the true gauge symmetries of the massless modes of the tower.  

In the case where spacetime is not flat, the notion of masslessness takes on greater richness.  There exist fields which are neither massless nor massive, but instead propagating a number of degrees of freedom intermediate between that of a massless and massive field.  These co-called ``partially massless" (PM) fields carry gauge invariances that remove some of the degrees of freedom of a generic massive mode, but the amount of gauge symmetry is smaller compared with that of a massless field, so we are left with a number of degrees of freedom greater than that of a massless field and less than that of a massive field~\cite{Deser:1983tm,Deser:1983mm,Higuchi:1986py,Brink:2000ag,Deser:2001pe,Deser:2001us,Deser:2001wx,Deser:2001xr,Zinoviev:2001dt,Skvortsov:2006at,Skvortsov:2009zu}.

The first non-trivial example of a partially massless field occurs in the case of spin 2 on de Sitter space.  At the special point $m^2={2\over D-1}\Lambda $ relating the mass and the cosmological constant, a massive spin-2 field acquires a scalar gauge invariance which removes its scalar helicity.  This case has been of interest due to a possible connection with the cosmological constant and cosmology (see ~\cite{deRham:2013wv} and the review \cite{Schmidt-May:2015vnx}), and there have been many studies of the properties of the theory and possible nonlinear extensions ~\cite{Zinoviev:2006im,Deser:2012qg,Hassan:2012gz,Hassan:2012rq,Hassan:2013pca,Deser:2013uy,Deser:2013bs,deRham:2013wv,Zinoviev:2014zka,Garcia-Saenz:2014cwa,Hinterbichler:2014xga,Joung:2014aba,Alexandrov:2014oda,Hassan:2015tba,Hinterbichler:2015nua,Cherney:2015jxp,Gwak:2015vfb,Gwak:2015jdo,Garcia-Saenz:2015mqi,Gwak:2016sma,Hinterbichler:2016fgl,Apolo:2016ort,Apolo:2016vkn}.

Here, we perform the KK reduction of a general massive spin-2 field, including the partially massless case.  We track the fate of the partially massless gauge symmetry as it undergoes dimensional reduction.  Like in the massless case, the higher KK harmonics of the gauge symmetry become algebraic St\"uckelberg-like symmetries associated with massive modes in the tower.  However, unlike in the massless case, the zero mode of the partially massless gauge symmetry does not become a lower-dimensional partially massless gauge symmetry.  Instead it serves to remove the zero mode scalar, the radion, and we are left with no partially massless graviton in the KK tower.  The radion is often a source of instability or an undesired 5th force and many mechanisms have been invented to stabilize or screen it, so a mechanism such as this to remove it completely via a gauge symmetry could be of interest.  The general pattern of fields which are present as a function of mass of the higher-dimensional field and curvature of the spacetime is quite intricate, as detailed in Section \ref{sec:spectrum}. 

For the linear KK reductions, we use the method advocated in \cite{Hinterbichler:2013kwa}.  The method works at the level of the action, keeps gauge invariance and covariance intact at all stages, and is applicable to any internal geometry.  The result is an action for the fields of the KK tower, from which the masses can be read off in terms of the eigenvalues of appropriate internal Laplacians.  In addition, it allows us to take into account all of the subtleties associated with zero modes, Killing vectors, conformal Killing vectors, etc. of the internal space.

We then move on to consider a fully non-linear example, the dimensional reduction of the ghost-free de Rham-Gabadadze-Tolley (dRGT) theory \cite{deRham:2010ik,deRham:2010kj} (reviewed in \cite{Hinterbichler:2011tt,deRham:2014zqa}) of a massive spin-2 field.  By keeping only the zero modes, to full non-linear order, we arrive at a massive scalar-vector-tensor theory that is a massive version of the original KK theory. By projecting out the vector we then obtain a family of massive scalar-tensor theories that are a special case of mass-varying massive gravity \cite{D'Amico:2011jj,Huang:2012pe} with varying $\alpha$'s \cite{Gumrukcuoglu:2013nza}.  Because these theories arise as consistent truncations of a ghost-free higher-dimensional theory, we expect them to be ghost free.

\textbf{Conventions and notation:} Our signature convention is mostly plus: $(-, +, +, \ldots)$. We work with a $D=(d+N)$-dimensional spacetime with indices $A, B, \ldots$ and coordinates $X^A$.  It is a product of a $d$-dimensional spacetime $\mathcal{M}$ and an $N$-dimensional space $\mathcal{N}$. $\mathcal{M}$ has spacetime indices $\mu, \nu, \ldots$ and coordinates $x^{\mu}$ and $\mathcal{N}$ has spatial indices $m, n, \ldots$ and coordinates $y^{m}$. 

\section{Setup}
\parskip=5pt
\normalsize

\subsection{Background Spacetime}

We begin by choosing a suitable direct product background metric on which the higher-dimensional massive graviton can propagate. Massless gravitons are only known to consistently propagate on background spacetimes that are Einstein spaces, i.e. solutions to the vacuum Einstein equations with a cosmological constant~\cite{Aragone:1979bm, Deser:2006sq}. Similarly, the only known backgrounds that partially massless spin-2 fields can propagate on are Einstein spaces \cite{Deser:2012qg,Gover:2014vxa}. There is no such restriction for general massive gravitons, which can propagate on an arbitrary background \cite{Bernard:2014bfa,Bernard:2015mkk,Bernard:2015uic}. In this paper we are most interested in the partially massless limits of massive gravity, so we will only consider massive gravitons on Einstein spaces.  It would be interesting to see what happens with these limits on non-Einstein backgrounds.

We therefore consider a $D=(d+N)$-dimensional background metric $G_{AB}$ that
satisfies the condition for an Einstein space, 
\be \label{ES}
R_{AB} = \frac{R_{(D)}}{D} G_{AB},
\ee
with constant scalar curvature ${R_{(D)}}$,
which is equivalent to demanding that the metric satisfies Einstein's equations 
\be
R_{AB}-\frac{1}{2} R_{(D)} G_{AB} + \Lambda_{(D)} G_{AB} =0
\ee
with cosmological constant
\be \label{curvature}
\quad \Lambda_{(D)} = \frac{(D-2)}{2D} R_{(D)}.
\ee
To perform a KK reduction we require the background to be in the form of a product space $\mathcal{M} \times \mathcal{N}$, so we take the background metric to be of the form
\be
G_{AB}(x,y) = 
\left( \begin{array}{cc}
g_{\mu \nu}(x) & 0 \\
0 & \gamma_{mn}(y)
\end{array} \right),
\ee
where $g_{\mu \nu}$ is the metric on the $d$-dimensional spacetime $\mathcal{M}$ with coordinates $x^{\mu}$ and $\gamma_{mn}$ is the metric on the $N$-dimensional space $\mathcal{N}$ with coordinates $y^{m}$. Writing \eqref{ES} in terms of the lower-dimensional metrics and their curvatures gives
\be
R_{\mu \nu} = \frac{R_{(d)}}{d} g_{\mu \nu}, \quad  R_{m n} = \frac{R_{(N)}}{N} \gamma_{m n},
\ee
where
\be
R_{(d)} = \frac{2d}{d+N-2}\Lambda_{(D)}, \quad R_{(N)} = \frac{2N}{d+N-2}\Lambda_{(D)},
\ee
which shows that both $\mathcal{M}$ and $\mathcal{N}$ must also be Einstein spaces. Note also that the curvatures satisfy
\be \label{equality}
\frac{R_{(D)}}{D} = \frac{R_{(d)}}{d} =\frac{R_{(N)}}{N}.
\ee

We further define the partially massless masses for each Einstein manifold with more than 2 dimensions:
\be
m^2_{(D) \rm{PM}} \equiv \frac{D-2}{D-1}\frac{R_{(D)}}{D}, \quad m^2_{(N) \rm{PM}} \equiv \frac{N-2}{N-1} \frac{R_{(N)}}{N}, \quad m^2_{(d) \rm{PM}} \equiv \frac{d-2}{d-1}\frac{R_{(d)}}{d}.
\ee
For $R<0$ these squared masses are all negative. By \eqref{equality} we have $m^2_{(D) \rm{PM}}> m^2_{(d) \rm{PM}}$ when $R>0$.

\subsection{Linear Action}

We now consider massive spin-2 perturbations $H_{AB}$ about the Einstein background $G_{AB}$. These are described by the Fierz-Pauli \cite{Fierz:1939ix} Lagrangian in $D$ dimensions,
\begin{align} \label{FPaction}
\frac{\mathcal{L}_{\rm{FP}, D}}{\sqrt{-G}}  = & -\frac{1}{2} \nabla_C H_{AB} \nabla^C H^{AB} + \nabla_C H_{AB} \nabla^B H^{AC} - \nabla_A H_{(D)} \nabla_B H^{AB} + \frac{1}{2} \nabla_{A} H_{(D)} \nabla^{A} H_{(D)} \nonumber \\
& + \frac{R_{(D)}}{D}\left( H_{AB} H^{AB} - \frac{1}{2} H^2_{(D)}\right)- \frac{1}{2} m^2 \left( H_{AB} H^{AB} -  H^2_{(D)}\right),
\end{align} 
where $\nabla_{A}$ is the background covariant derivative, $R_{(D)}$ is the constant curvature given by \eqref{curvature}, and $m$ is the graviton mass. The Lagrangian \eqref{FPaction} describes different numbers of propagating degrees of freedom depending on the value of the mass and scalar curvature. We will consider only $m^2>0$; the general dimensional reduction for $m=0$ was worked out in \cite{Hinterbichler:2013kwa}.  When $R>0$, the Higuchi bound states that a spin-2 field is only stable if $m \geq m_{(D)\rm{PM}}$. The different cases we consider are: 
\begin{enumerate}
\item $0 <m < m_{(D) \rm{PM}}$, $R>0$. This describes an unstable massive graviton on a positively curved Einstein background with  $(D+1)(D-2)/2$ degrees of freedom, one of which is a ghost-like degree of freedom.
\item $m = m_{(D) \rm{PM}}$, $R>0$. This describes a stable partially massless graviton on a positively curved Einstein background which has the scalar gauge symmetry
\be \label{PMsymmetry}
\delta H_{AB} = \nabla_A \nabla_B \alpha + G_{AB} \frac{R_{(D)}}{D(D-1)} \alpha  
\ee
and propagates  $(D+1)(D-2)/2-1$ degrees of freedom. 
\item $m > m_{(D) \rm{PM}}$. This describes a stable massive spin-2 particle  on an Einstein background  with $(D+1)(D-2)/2$ degrees of freedom. This includes the case $R \leq 0$. 
\end{enumerate}
There are other values of $m^2$ that are special from the perspective of lower-dimensional physics and we discuss these later.

\subsection{Symmetric Tensor Decomposition}

In order to find the $d$-dimensional KK spectrum we must decompose our higher-dimensional field into orthonormal eigenfunctions of various Laplacian operators defined on the internal manifold $\mathcal{N}$. This is the generalization to arbitrary manifolds of the familiar Fourier decomposition used in the original KK construction and allows us to integrate out the dependence of fields on the internal space $\mathcal{N}$. The field $H_{AB}$ breaks up into pieces $H_{\mu \nu}$, $H_{\mu n}$, and $H_{mn}$, which transform respectively as a scalar, vector and symmetric tensor on the internal manifold. We thus need to decompose general scalar, vector and symmetric tensor fields into eigenmodes of appropriate Laplacian operators on an arbitrary compact manifold. The appropriate decomposition is given by the Hodge decomposition for scalar and vector fields and its generalization to symmetric tensor fields \cite{besse2007einstein,Ishibashi:2004wx}, which for the different components of $H_{AB}$ give
\begin{align}
H_{\mu \nu} =&  \sum_a h_{\mu \nu}^a \psi_a + \frac{1}{\sqrt{\mathcal{V}_n}} h_{\mu \nu}^0\, ,\nn \\
H_{\mu n} = & \sum_i A_{\mu}^i Y_{n, i}+\sum_a A_{\mu}^a \nabla_n \psi_a \, ,\nn\\
H_{mn} = & \sum_{\mathcal{I}} \phi^{\mathcal{I}} h_{mn, \mathcal{I}}^{TT} + \sum_{i \neq \rm{Killing}} \phi^i \left( \nabla_m Y_{n,i} + \nabla_n Y_{m,i} \right)\, , \nn\\
& + \sum_{a \neq \rm{conformal}} \phi_a \left( \nabla_m \nabla_n \psi_a -  \frac{1}{N} \nabla^2 \psi_a \gamma_{mn} \right)+ \sum_a \frac{1}{N} \bar{\phi}^a \psi_a \gamma_{mn} +\frac{1}{N} \frac{1}{\sqrt{\mathcal{V}_N}} \phi_0 \gamma_{mn}. \label{hmnsum3}
\end{align}
The fields $\psi_a$ are a basis of positive orthonormal eigenmodes of the scalar Laplacian on $\mathcal{N}$ with positive eigenvalue $\lambda_a$ (i.e. excluding the zero mode),
\be
\int d^N y \sqrt{\gamma} \psi^{a*}(y) \psi_b(y) = \delta^a_b,
\ee
\be
\left( \Box_{(y)} + \lambda_a \right) \psi^a(y)  =0, \quad \lambda_a > 0,
\ee
and $\mathcal{V}_N$ is the volume of $\mathcal{N}$
\be
\mathcal{V}_N \equiv \int d^N y \sqrt{\gamma}. 
\ee
The fields $Y_{n, i}$ are a basis of orthonormal transverse eigenvectors of the vector Laplacian on $\mathcal{N}$ with eigenvalues $\lambda_i$,
\be
\int d^N y \sqrt{\gamma} Y^*_{i,m}(y) Y^m_j(y) = \delta_{ij},
\ee
\be
\nabla^n_{(y)} Y_{i,n}(y) =0,
\ee
\be
\Delta Y_{i,n} \equiv - \Box_{(y)} Y_{i,n}(y) + R_n^{\; m} Y_{i,m}(y) = \lambda_i Y_{i,n}(y), \quad \lambda_i \geq 0.
\ee
 The fields $h_{mn, \mathcal{I}}^{TT} $ are a basis of symmetric transverse traceless orthonormal eigenmodes of the Lichnerowicz operator \cite{Lichnerowiczlaplacian} on $\mathcal{N}$ with eigenvalues $\lambda_{\mathcal{I}}$,
 \be
\nabla^m_{} h_{mn, \mathcal{I}}^{TT}(y) = \gamma^{m n} h_{mn, \mathcal{I}}^{TT}(y) =0,
\ee
\be
\int d^N y \sqrt{\gamma} h_{mn}^{TT, \mathcal{I} *}(y) h^{mn, TT}_{\mathcal{J}}(y) = \delta^{\mathcal{I}}_{\mathcal{J}},
\ee
\be
\Delta_{L, (y)} h_{mn, \mathcal{I}}^{TT}(y) \equiv -\Box_{(y)} h_{mn, \mathcal{I}}^{TT}(y) + \frac{2 R_{(N)}}{N} h_{mn, \mathcal{I}}^{TT}(y) - 2 R_{mpnq} h^{pq \; TT}_{\mathcal{I}}(y) = \lambda_{\mathcal{I}} h_{mn, \mathcal{I}}^{TT}.
\ee
The bases of eigenmodes can always be chosen to be real functions, however it is often convenient to allow them to be complex. Ensuring that the physical fields are real then requires a reality condition, e.g,  restricting the $\phi^a(x)$ to satisfy 
$
\phi^{a*}(x)= \phi^{\bar{a}}(x),
$
where $\bar{a}$ denotes some involution on the set of indices $a$. 
The sum over $i$ in the decomposition of $H_{mn}$ in \eqref{hmnsum3} excludes Killing vectors, which satisfy
\be
\nabla_m Y_{n,i}(y) + \nabla_n Y_{m,i}(y) =0.
\ee
Nontrivial Killing vectors can exist on closed Einstein manifolds only when $R_{(N)} \geq 0$ and they are precisely the transverse eigenmodes of the vector Laplacian with eigenvalue $\lambda_i = 2R_{(N)}/N$, which is the smallest possible eigenvalue for transverse eigenmodes of the vector Laplacian on a closed Einstein manifold with nonnegative curvature. Similarly, the sum over $a$ excludes conformal scalars, which satisfy
\be
\nabla_m \nabla_n \psi_a(y) -  \frac{1}{N} \Box_{(y)} \psi_a(y) \gamma_{mn} =0.
\ee
Conformal scalars exist only on manifolds that are isometric to the sphere \cite{obata1962} and are precisely the eigenmodes of the scalar Laplacian with eigenvalue 
\be \label{obata}
\lambda_{a=\rm{conformal} }=\frac{R_{(N)}}{N-1}.
\ee
By the Lichnerowicz bound \cite{lichnerowiczbound}, \eqref{obata} is the smallest possible eigenvalue for eigenmodes of the scalar Laplacian on a closed Einstein manifold with positive curvature. For more details about the decomposition of $H_{AB}$ and why it takes the form it does, we refer to \cite{Hinterbichler:2013kwa}. 

\subsection{Reduction of the PM Gauge Transformations}
When $m = m_{(D) \rm{PM}}$ the $D$-dimensional Lagrangian acquires the partially massless gauge symmetry \eqref{PMsymmetry}. To find how this gauge symmetry acts on the lower-dimensional fields we first decompose the partially massless gauge parameter in eigenmodes of the internal Laplacian,
\be
\alpha= \sum_a \alpha^a \psi_a + \frac{1}{\sqrt{\mathcal{V}_N}} \alpha_0.\label{gaugeexppm}
\ee
Expanding the gauge transformation \eqref{PMsymmetry} using \eqref{hmnsum3} and \eqref{gaugeexppm} and equating components then gives the following gauge transformations for the lower-dimensional fields:
\begin{align}
\delta h_{\mu \nu}^a = & \nabla_{\mu} \nabla_{\nu} \alpha^a + g_{\mu \nu} \frac{R_{(D)}}{D(D-1)} \alpha^a, & \delta A_{\mu}^a = & \nabla_{\mu} \alpha^a,\nn \\
\delta h_{\mu \nu}^0 = & \nabla_{\mu} \nabla_{\nu} \alpha^0 + g_{\mu \nu} \frac{R_{(D)}}{D(D-1)} \alpha^0, & \delta A_{\mu}^i = & 0\, ,  \nn\\
 \delta \bar{\phi}^a = & \left( \frac{N R_{(D)}}{D(D-1)} - \lambda_a \right) \alpha^a,  & \delta \phi^a = & \alpha^a, \quad a \neq \rm{conformal} \nn\\
\delta \phi^0  = & \frac{N R_{(D)}}{D(D-1)} \alpha^0, & \delta \phi^i = & 0, \quad i\neq \rm{Killing} \nn\\
 \delta \phi^{\mathcal{I}} =&  0\, . & 
\end{align}

We can see that the higher-dimensional partially massless symmetry acts as a St\"uckelberg symmetry on some of the lower-dimensional fields. In particular, we can use $\alpha^0$ and $\alpha^a$ to gauge fix 
\begin{equation} \label{PMgaugefix}
\phi^0=\phi^{a \neq \rm{conformal}}=\bar{\phi}^{a = \rm{conformal}}=0.
\end{equation}
This completely fixes the gauge symmetry, which indicates already that there will be no partially massless fields in the lower-dimensional theory. Note that some of the gauge symmetry would remain if 
\be
\lambda_{a = \rm{conformal}}= \frac{N R_{(D)}}{D(D-1)} ,
\ee 
but this is forbidden by \eqref{equality} and \eqref{obata} for $D>N$.  

\section{Dimensionally Reduced Action}

Plugging the decomposition \eqref{hmnsum3} into \eqref{FPaction} and integrating over the extra dimensions using orthogonality gives the $d$-dimensional action
\be \label{Ltot}
S= \int d^d x \sqrt{-g} (\mathcal{L}_{\mathcal{I}}+\mathcal{L}_i +\mathcal{L}_0 + \mathcal{L}_a  ), 
\ee
where the $\mathcal{L}$'s are quadratic in the $d$-dimensional fields appearing in the decomposition of $H_{AB}$. The different terms in \eqref{Ltot} are given by:
\begin{align}
\mathcal{L}_{\mathcal{I}} = & \sum_{\mathcal{I}} -\frac{1}{2}|\partial \phi^{\mathcal{I}}|^2 - \frac{1}{2} \left( \lambda_{\mathcal{I}}+m^2-2\frac{R_{(d)}}{d} \right) | \phi^{\mathcal{I}}|^2\, , \nn\\
\mathcal{L}_i = & \sum_{i} -\frac{1}{2}|F_{\mu \nu}^i|^2 - \left( \lambda_i - \frac{2 R_{(d)}}{d} \right)|A_{\mu}^i - \partial_{\mu} \phi^i|^2 - m^2|A_{\mu}^i|^2 -\frac{1}{2}m^2 \left(\lambda_i - \frac{2R_{(d)}}{d}\right)|\phi^i|^2 \, ,\nn\\ 
\mathcal{L}_0 = & \varepsilon(h^0)_{m^2} + h^{\mu \nu, 0}\left( \nabla_{\mu} \nabla_{\nu} \phi^0 - \Box \phi^0 g_{\mu \nu} \right) + h^0 \phi^0 \left( m^2 - \frac{R_{(d)}}{d} \right) \nn\\ 
&+ \frac{N-1}{2N} \left[ \left(\partial \phi^0\right)^2 + (\phi^0)^2 \left(m^2 - m^2_{(N) \rm{PM}} \right) \right] \, ,\nn\\
\mathcal{L}_a = & \sum_{a } \epsilon({h}^a)_{m^2 + \lambda_a} -\frac{1}{2} \lambda_a |{F}_{\mu \nu}^a|^2 +  \lambda_a \left(\frac{2 R_{(d)}}{d} -m^2\right)|{A}_{\mu}^a|^2  + \frac{N-1}{2N} \left( |\partial_{\mu} \bar{\phi}^a|^2 +m^2  |{\bar{\phi}}^a|^2 \right) \nn\\
&+ \frac{1}{2} \Big[   2 \lambda_a {h}^{\mu \nu, a *}(\nabla_{\mu}A_{\nu}^a-g_{\mu\nu}\nabla^{\alpha}A_{\alpha}^a)   +{h}^{\mu \nu, a *}(\nabla_{\mu}\nabla_{\nu} \bar{\phi}^a - g_{\mu \nu} \Box \bar{\phi}^a ) \nn\\
&  + {h}^{*a} \bar{\phi}^a \left(\frac{N-1}{N}\lambda_a+m^2-\frac{R_{(d)}}{d} \right)-2\lambda_a\frac{N-1}{N} \nabla^{\mu}{A}_{\mu}^{a*} \bar{\phi}^a +\rm{c.c}\Big]\nn\\
& + \frac{N-1}{N}\lambda_a\left(\lambda_a-\frac{R_{(N)}}{N-1} \right)\left[  \frac{N-2}{2N\lambda_a}| \lambda_a \phi^a + \bar{\phi}^a|^2-\frac{1}{2} |\partial \phi^a|^2 - \frac{1}{2} m^2 |\phi^a|^2 \right. \nn\\
& +((h^a/2  -\nabla_{\mu}A^{\mu}_a) \phi^{a*} + \rm{c.c}) \bigg],
\end{align}
where we have defined $\epsilon(h)_{m^2}$ as the $d$-dimensional massive graviton action, i.e. the $d$-dimensional version of \eqref{FPaction}.

\subsection{Diagonalization}  \label{sec:diagonalization}

To determine the spectrum and stability of $S$ we need to diagonalize $\mathcal{L}_i$, $\mathcal{L}_0$ and $\mathcal{L}_a$. 
The required demixing transformations depend on $m^2$ and details of the spaces involved. 
Here we present the final diagonalized Lagrangians $\mathcal{L}_i$, $\mathcal{L}_0$ and $\mathcal{L}_a$, leaving the details of the required transformations to the appendix. The fields in the diagonalized Lagrangians are combinations of the fields in the original Lagrangian but to avoid clutter we give them the same names. We define
\be \label{massdefinition}
m^2_k \equiv m^2+ \lambda_k - \frac{2R_{(d)}}{d},
\ee
where $k= \mathcal{I},$ $0,$ $i$, or $a$ and $\lambda_0=0$.

\subsubsection*{Diagonalized $\mathcal{L}_i$}
The diagonalized Lagrangian $\mathcal{L}_i$ is given by
\begin{align}
\mathcal{L}_i & = \sum_{i \neq \rm{Killing}} -\frac{1}{2}|F_{\mu \nu}^{i}|^2 - m_i^2 |A_{\mu}^{i}|^2 - \frac{m^2 \left(m^2_i-m^2\right)}{m_i^2} \left( |\partial \phi^i|^2+\frac{1}{2} m_i^2|\phi^i|^2 \right) \nn\\ & + \sum_{i = \rm{Killing}} -\frac{1}{2}|F_{\mu \nu}^i|^2 - m^2|A_{\mu}^i|^2.
\end{align}

\subsubsection*{Diagonalized $\mathcal{L}_0$}
When $m \neq m_{(d) \rm{PM}}$ and $d>1$ the diagonalized $\mathcal{L}_0$ is
\be \label{L0demixed}
\mathcal{L}_0 =  \epsilon(h^0)_{m^2}- \frac{D-1}{2N(d-1)}\frac{m^2-m^2_{(D)\rm{PM}}}{m^2-m^2_{(d)\rm{PM}}}\left((\partial \phi^0)^2 +m^2_0 (\phi^0)^2 \right).
\ee
When $m = m_{(d) \rm{PM}}$ 
\be \label{PMzeromode1}
\mathcal{L}_0 = \epsilon({h}^0)_{m^2_{(d)\rm{PM}}} - \frac{1}{4} m^2_{(d)\rm{PM}} {h}_0^2. 
\ee
When $d=1$ many of the terms vanish and $\mathcal{L}_0$ reduces to a nondynamical Lagrangian. The Lagrangian \eqref{PMzeromode1} is unusual, and we will say more about this below.

\subsubsection*{Diagonalized $\mathcal{L}_{a}$}

When $m_a \neq 0$, $N>1$, and $d>1$ the Lagrangian $\mathcal{L}_{a}$ is
\begin{align}  
\label{diaglanoncon}
\mathcal{L}_a =&  \sum_{a } \epsilon({h}^{a})_{m^2 + \lambda_a} -\frac{m^2 \lambda_a}{m^2+\lambda_a}\left( \frac{1}{2} |{F}_{\mu \nu}^{a}|^2 + m_a^2 |{A}_{\mu}^{a}|^2\right)\nn\\ & - \sigma_a \left( | \partial {\phi}^a |^2 + m_a^2 |{\phi}^a|^2 \right) - \bar{\sigma}_a \left( | \partial \bar{\phi}^a |^2 + m_a^2 |\bar{\phi}^a|^2 \right) ,
\end{align}
where $\sigma_a$ and $\bar{\sigma}_a$ are functions of $\lambda_a$, $R$, $d$ and $N$ that are defined in the appendix -- in particular, $\sigma_{a = \rm{conformal}}=0$.
When  $m_a = 0$ we can write $\mathcal{L}_{a}$ as
\begin{align} \label{conformspecial2}
 \mathcal{L}_{a } = & \sum_{m^2_a=0 } \epsilon({h}^{a})_{2 R_{(d)}/d} -\frac{m^2 \lambda_a}{m^2+\lambda_a} \left( |\partial_{\mu} A^{a}_{\nu}|^2 +\frac{R_{(d)}}{d} |A_{\mu}^a |^2 \right)\nn \\
& -\frac{(N-1) (D-1)\left( m^2- m^2_{(N) \rm{PM}} \right)\left( m^2- m^2_{(D) \rm{PM}} \right)}{2d} | \partial{\phi}^{a} |^2 .
\end{align}
Achieving $m_a=0$ requires tuning $m$ and $\lambda_a$ but can occur for $m \leq m_{(N)\rm{PM}}$.
When $N=1$ all scalars are conformal and for $d>1$ we get 
\begin{align} \label{conformalN1}
\mathcal{L}_a = & \sum_{a = \rm{conformal}} \epsilon({h}^{a})_{m_a^2}-\frac{m^2 \lambda_a}{ m^2_a}\left( \frac{1}{2} |{F}_{\mu \nu}^{a}|^2 + m^2_a |{A}_{\mu}^{a}|^2\right) - \frac{d m^4 }{2(d-1)m^4_a} \left( \left| \partial \bar{\phi}^a \right|^2 +  m^2_a |\bar{\phi}^a|^2 \right).
\end{align}
Lastly, when $d=1$ we get
\begin{align} \label{noncond1}
\mathcal{L}_a = &  -\frac{N-1}{2N} \sum_{a\neq \rm{conformal}} \left( |\partial_{\mu} \bar{\phi}^a|^2 +m^2_a |{\bar{\phi}}^a|^2 \right).
\end{align}

Let us now comment on the degrees of freedom described by \eqref{PMzeromode1}. The $|{h}_0|^2$ term spoils the PM gauge symmetry and the Fierz-Pauli tuning of the mass term, so we expect that \eqref{PMzeromode1} describes an extra ghostly scalar degree of freedom\footnote{In the case of linearized general relativity, adding a term $\beta h^2$ spoils the linearized diffeomorphism symmetry but gives a theory that is the same as linearized general relativity up to global degrees of freedom.}.  A $3+1$ analysis reveals that for $d=4$ \eqref{PMzeromode1} describes four PM modes in addition to two scalar modes, one of which is a ghost\footnote{Another way to see the ghost once we know that there are six degrees of freedom is to perform the St\"uckelberg transformation
\be \label{pmstuck}
{h}_{\mu \nu}^0 \rightarrow {h}_{\mu \nu}^0 + \nabla_{\mu} \nabla_{\nu} \varphi + g_{\mu \nu} \frac{m^2_{(d)\rm{PM}} }{(d-2)} \varphi,
\ee
which introduces a new scalar field $\varphi$. This transformation has the form of a PM transformation so only affects the $|{h}_0|^2$ term. The resultant Lagrangian is then invariant under the PM St\"uckelberg symmetry
\be
\delta h_{\mu \nu} =  \nabla_{\mu} \nabla_{\nu} \alpha + g_{\mu \nu} \frac{m^2_{(d)\rm{PM}} }{(d-2)} \alpha, \quad \delta \varphi = - \alpha. 
\ee
Now rescale $\varphi \rightarrow \varphi/m$ and take the decoupling limit $m\rightarrow 0$ while preserving the PM relation $m = m_{(d)\rm{PM}}$. In this limit the partially massless kinetic term breaks up into massless tensor and vector modes and the surviving scalar term is higher derivative $\sim \varphi \Box^2 \varphi$, which shows that there is a scalar ghost.}.  In fact, it is not possible to transform \eqref{PMzeromode1} to a set of decoupled kinetic terms of familiar fields.  It is inherently undemixable, and describes a new kind of irreducible representation which is a combination of a spin-2 field and a spin-0 field on curved space; it is a field theoretic manifestation of a so-called ``extended module" which mixes spin-0 and spin-2 fields, seen in the Hilbert space of the non-unitary $\square^2$ conformal field theory in three dimensions \cite{Brust:2016gjy}.  An equivalent form of \eqref{PMzeromode1} occurs in the holographic dual of the $\square^2$ conformal field theory \cite{Brust:2016zns}.

\subsection{Spectrum and Stability} \label{sec:spectrum}

Now that we have diagonalized the action we can determine its spectrum and stability. The spectrums of $\mathcal{L}_{\mathcal{I}}$ and $\mathcal{L}_i$ are straightforwardly determined, so we discuss these first. Next we discuss the spectrums of $\mathcal{L}_0$ and $\mathcal{L}_a$ for the three cases where $m$ is less than, equal to, or larger than $m_{(D) \rm{PM}}$. We assume that $d>1$ and $N>1$ until the end where we discuss the special cases $N=1$ and $d=1$. We conclude with a summary of these stability results.

\subsubsection*{Spectrum of $\mathcal{L}_{\mathcal{I}}$}
The Lagrangian $\mathcal{L}_{\mathcal{I}}$ describes a tower of massive scalars with squared masses given by $m_{\mathcal{I}}^2$, where $m_{\mathcal{I}}^2$ is defined by \eqref{massdefinition}.
The stability of a scalar mass term depends on the background curvature. For a flat or de Sitter space stability requires $m_{\mathcal{I}}^2 \geq 0$ and for an anti-de Sitter space stability requires
\be
m_{\mathcal{I}}^2 \geq \frac{d-1}{4} \frac{R_{(d)}}{d},
\ee
which is the Breitenlohner-Friedman bound \cite{Breitenlohner:1982bm,Breitenlohner:1982jf}. We do not know of bounds for the more general case of a scalar on an Einstein space, so we take these bounds for maximally symmetric spaces as our conditions for stability. There exist positively curved compact manifolds with large negative eigenvalues of the Lichnerowicz operator (such as the B\"ohm metrics on products of spheres \cite{Gibbons:2002th}). This means that the scalars $\phi^{\mathcal{I}}$ can be tachyonic, but for any given internal space $\mathcal{N}$ there is some smallest $m$ that stabilizes them. 

\subsubsection*{Spectrum of $\mathcal{L}_i$}

The spectrum of $\mathcal{L}_i$ is a tower of massive vectors with mass squared $m_i^2$ and a tower of massive scalars with mass squared  $m_i^2/2$, one for each non-Killing transverse vector of the internal space, and a massive vector with mass $m$ for each Killing vector of the internal space. For a non-Killing vector we have $\lambda_i > 2 R_{(N)}/N$ and hence $m_i^2>m^2>0$.
 Stability of vectors requires only that their mass be positive and hence all fields are stable in this sector.  
 
\subsubsection*{Spectrum of $\mathcal{L}_0$ and $\mathcal{L}_a$ for $0<m<m_{(D) \rm{PM}}$}

We first consider the spectrum of $\mathcal{L}_0$ and $\mathcal{L}_a$ for the mass range $0<m<m_{(D) \rm{PM}},$ which is relevant for $D>2$ and $R>0$. This corresponds to a higher-dimensional theory with an unstable graviton, since its mass falls below the Higuchi bound, so we expect instabilities in the lower-dimensional theory. The spectrum consists of a zero-mode massive scalar and a zero-mode massive graviton, plus a massive scalar, vector and tensor for each conformal scalar. There is also a tower of massive gravitons, a tower of massive vectors, and two towers of massive scalars, one field in each of the towers for each non-conformal scalar in the internal space. Below we discuss the stability of this spectrum for each spin.

 \underline{Spin 2:} The zero-mode graviton has mass $m$ and thus by the Higuchi bound is unstable for  $m<m_{(d) \rm{PM}}$ and stable for $m>m_{(d) \rm{PM}}$. When $m = m_{(d) \rm{PM}}$ the zero-mode sector \eqref{PMzeromode1} contains a new kind of irreducible representation which can be thought of as an undemixable combination of a massive graviton and a ghost, as discussed further at the end of Section \ref{sec:diagonalization} and in the appendix. The other gravitons have mass squared $m^2+\lambda_a>0$ and are always stable. 

 \underline{Spin 1:} For $m_a \neq 0$ the massive vectors have mass $m_a$, defined in \eqref{massdefinition}, so their stability depends on $m$ and $\lambda_a$. For $m<m_{(N) \rm{PM}}$, the massive vectors corresponding to conformal scalars are tachyonic since 
\be
m_{a=\rm{conformal}}^2=m^2-m^2_{(N) \rm{PM}},
\ee
and the massive vectors corresponding to non-conformal scalars can be tachyonic or stable depending on the sign of $m_a^2$.  For $m>m_{(N) \rm{PM}}$ the massive vectors are all stable. When $m_{a}=0$ the spatial components of the vector field describe $d-1$ scalars with masses squared equal to $R_{(d)}/d$, which violate the Breitenlohner-Friedman stability bound when $R<0$ and $d<5$, and the temporal component of the vector field is a scalar ghost. 

\begin{figure}[h!]
\begin{center}
\epsfig{file=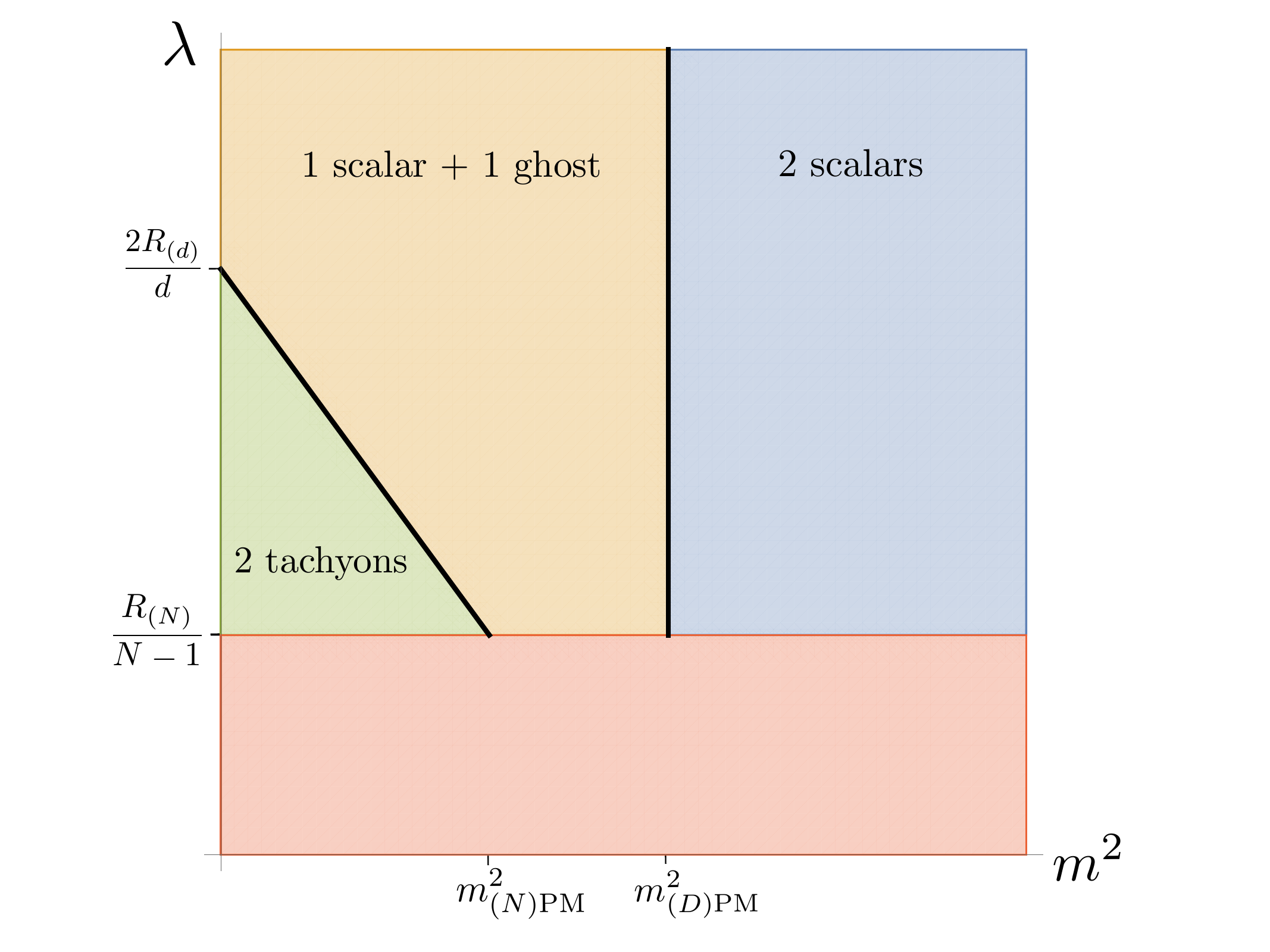,height=4in,width=5.0in}
\caption{Plot of the field content of the nonconformal scalar sector in the $(m^2, \lambda)$ plane for $R>0$, $d>1$, $N>2$. The lower rectangle corresponds to eigenvalues at or below the Lichnerowicz bound. The rightmost region has 2 healthy scalars, the adjacent region has 1 healthy scalar and 1 ghost, and the triangular region has 2 tachyons. The angled line is $m_a=0$ and the vertical line is $m=m_{(D)\rm{PM}}$, which has 1 healthy scalar.  }
\label{fig:nonconformal}
\end{center}
\end{figure}

 \underline{Spin 0:} The zero-mode scalar -- the radion mode -- is tachyonic for $m<m_{(d) \rm{PM}}$ and ghostly for $m_{(d) \rm{PM}}<m<m_{(D) \rm{PM}}$. When $m=m_{(d) \rm{PM}}$ the zero-mode scalar gets folded into the graviton field giving \eqref{PMzeromode1}.
The scalars coming from the conformal modes are tachyonic for $m<m_{(N) \rm{PM}}$ and ghostly for $m_{(N) \rm{PM}}<m<m_{(D) \rm{PM}}$. The scalars in the two towers, corresponding to non-conformal scalars in the internal space, are both tachyonic if $m^2_a<0$ and have one healthy scalar and one ghost if $m^2_a>0$. When $m_a=0$ there is one massless scalar if $a\neq \rm{conformal}$ and no scalars if $a=\rm{conformal}$, since then $m=m_{(N) \rm{PM}}$. The scalar spectrum coming from the nonconformal modes of the internal space is summarized in Figure \ref{fig:nonconformal}.

\subsubsection*{Spectrum of $\mathcal{L}_0$ and $\mathcal{L}_a$ for $m=m_{(D) \rm{PM}}$}
Now consider the case when the higher-dimensional spin-2 field is partially massless, $m=m_{(D)\rm{PM}}$. The spectrum consists of a tower of massive gravitons, a tower of massive vectors, and a tower of massive scalars. There is also a massive graviton and massive vector for each conformal mode, and a zero-mode massive graviton. All of these fields are stable. The effect of the higher-dimensional partially massless gauge symmetry is to remove a tower of scalar fields, including the would-be unstable radion, which is consistent with the gauge fixing \eqref{PMgaugefix}. There are no partially massless fields in the lower-dimensional spectrum since the lightest massive graviton has mass $m_{(D) \rm{PM}}$ and this is larger than $m_{(d) \rm{PM}}$. 

\subsubsection*{Spectrum of $\mathcal{L}_0$ and $\mathcal{L}_a$ for $m_{(D) \rm{PM}}<m$}
Lastly, consider the case of a stable higher-dimensional massive graviton, $m> m_{(D)\rm{PM}}$. This includes all cases when $R \leq0$. The field content is the same as for $m<m_{(D) \rm{PM}}$. The zero-mode scalar is tachyonic for $m^2_{(D)\rm{PM}} < m^2 < 2 R_{(d)}/d$ and nontachyonic for $m^2> 2 R_{(d)}/d$. The scalar is massless when $m^2 = 2 R_{(d)}/d$. All the other fields are stable. 

\subsubsection*{Spectrum When $d=1$ or $N=1$} \label{sec:N1}
When $N=1$, i.e. when $\mathcal{N}$ is the circle $S^1$ with radius $r$, there are no symmetric transverse traceless tensors on $\mathcal{N}$ so $\mathcal{L}_{\mathcal{I}}$ is zero. There are no non-Killing transverse vectors and the only Killing vector is the constant vector on the circle, so $\mathcal{L}_i$ contains a single vector with mass $m$, which is a massive version of the original KK photon. There is a zero-mode massive graviton and a zero-mode massive scalar, both with mass $m$. There is also a tower of massive gravitons, vectors and scalars all with masses squared
\be
m_a^2 = m^2+ \frac{a}{r^2},
\ee
where $a=1, 2, 3, \ldots$ When $d=1$ there are massive scalars for each Lichnerowicz mode, each non-Killing mode and each non-conformal mode. 

\subsubsection*{Summary of Stability Results}

We have found that avoiding the radion instability requires either $m^2 \geq 2 R_{(D)}/D$ or $m^2 = m^2_{(D) \rm{PM}}$. The partially massless value thus corresponds to an isolated line of stability in the $(m^2, R)$ plane since the PM gauge symmetry renders the unstable radion pure gauge. There may be other instabilities depending on the size of $m^2$ and the spectrum of eigenvalues of the Lichnerowicz operator on the internal space.

\section{Interactions}

So far we have been working with quadratic actions only. Including interactions requires a nonlinear $D$-dimensional theory of massive gravity. Adding a generic interaction term to the linear massive Fierz-Pauli action \eqref{FPaction} gives a theory with a massive graviton and an additional scalar ghost, the Boulware-Deser ghost \cite{Boulware:1973my}, which becomes active at the non-linear level.  Our previous linear analysis is insensitive to the presence of a Boulware-Deser ghost in the higher-dimensional theory. However, if we restrict to higher-dimensional theories with the ghost-free dRGT \cite{deRham:2010kj} interactions, then the quadratic action describes all propagating modes and our previous results determine the full KK spectrum\footnote{dRGT massive gravity itself can be derived from ``dimensional deconstruction" of Einstein gravity \cite{ArkaniHamed:2002sp,ArkaniHamed:2003vb,deRham:2013awa}, which can be thought of as a kind of discrete KK reduction.}.
A straightforward self-interacting theory of a single partially massless spin-2 field does not appear to exist~\cite{Deser:2013uy,deRham:2013wv,Joung:2014aba,Zinoviev:2014zka}, so we will not be able to address this case non-linearly.

It is possible, if tedious, to include interactions in the lower-dimensional theory by plugging the decomposition of $H_{AB}$ order-by-order into the higher-dimensional interactions and integrating over the resulting products of eigenmodes. Doing this to all orders gives a lower-dimensional re-writing of the higher-dimensional theory. For example, in the original KK setup, $D=5$ general relativity with a single compact extra dimension gives a $d=4$ theory with a massless graviton, vector and scalar, all interacting with an infinite tower of massive gravitons. Similarly, based on the results of the previous section, reducing $D=5$ dRGT massive gravity should give a $d=4$ theory of an infinite number of massive gravitons, vectors, and scalars, all interacting with one another.\footnote{
What is the cutoff of this theory? The strong coupling scale in $(n+4)$-dimensional massive gravity is given by
$
\Lambda_{\frac{n+6}{n+2}} \equiv \left(m^{\frac{4}{n+2}} M_{(n+4)} \right)^{\frac{n+2}{n+6}},
$
where $M_{(n+4)}$ is the $(n+4)$-dimensional Planck mass. The $d=4$ Planck mass is given by
$
M_{(4)}^2 \sim  M_{(n+4)}^{n+2} R^n,
$
where $R$ is the characteristic length scale of the internal space. This gives
\begin{equation} \label{cutoff}
\Lambda_{\frac{n+6}{n+2}} \sim \left( \frac{\Lambda_3^6}{R^n} \right)^{\frac{1}{n+6}},
\end{equation}
where $\Lambda_3 \equiv \left( m^2 M_{(4)} \right)^{\frac{1}{3}}$ is the $d=4$ strong coupling scale of massive gravity. Consistency of the effective field theory requires that the inverse size of the internal space is smaller than the cutoff, $R^{-1} < \Lambda_{\frac{n+6}{n+2}}$. Together with \eqref{cutoff} this implies that $\Lambda_{\frac{n+6}{n+2}} < \Lambda_3$, so dimensionally reducing in this way cannot raise the strong coupling scale above $\Lambda_3$.  
}

One tractable way to study nonlinear interactions is to consider a single compact extra dimension and to truncate to the zero modes. This truncation is consistent because translations in the extra dimension act as a global $U(1)$ symmetry on the four-dimensional fields, and the zero modes (which are constants on the circle) are the only singlets under this $U(1)$. 

Let us briefly review how this procedure works in the case of general relativity before applying it to massive gravity. We take the five-dimensional spacetime to be $M^4 \times S^1$, the product of four-dimensional Minkowski and a circle of length $L$. The five-dimensional Einstein-Hilbert action is 
\be
S = \frac{M_{(5)}^3}{2} \int d^4 x \int_0^L dy \sqrt{-G} R(G),
\ee
where $G_{AB}$ is the higher-dimensional metric and $M_{(5)}$ is the five-dimensional Planck mass. The zero mode truncation is accomplished by substituting the following ansatz for the metric
\begin{equation} \label{KKansatz}
G_{A B}(x) = \left( 
\begin{array}{ccc}
g_{\mu \nu}(x) + \phi^2(x) A_{\mu}(x) A_{\nu}(x) & \phi^2(x) A_{\mu}(x) \\
 \phi^2(x) A_{\nu}(x) & \phi^2(x)
\end{array} \right).
\end{equation}
Substituting into the action gives
\be
S = \frac{M_{(5)}^3}{2} L \int d^4 x \sqrt{-g} \phi \left( R - \frac{1}{4} \phi^2 F_{\mu \nu}^2  \right).
\ee
We can go to the Einstein frame and canonically normalize the scalar by defining $\tilde{g}_{\mu \nu} \equiv \phi g_{\mu \nu}$ and $\phi \equiv e^{- \psi}$. This gives
\be
S = \frac{M_{(4)}^2}{2}  \int d^4 x \sqrt{-\tilde{g}}  \left( \tilde{R} - \frac{1}{4} e^{- 3\psi} F_{\mu \nu}^2 -\frac{3}{2} (\partial \psi)^2 \right),
\ee
where the 4D Planck mass is given by $M_{(4)}^2 = M_{(5)}^3 L$.  
This describes an interacting massless graviton, vector and scalar. 

Now consider five-dimensional dRGT massive gravity on the spacetime $M^4 \times S^1$. The action is given by
\be \label{5ddRGT}
S = \frac{M_{(5)}^3}{2} \int d^4 x \int_0^L dy \sqrt{-G} \left( R(G) - \frac{m^2}{4} \sum_{n=0}^5 \beta_n S_n \left( \sqrt{G^{-1} \eta} \right) \right),
\ee
where $\eta_{M N}$ is the 5D fiducial metric, which we take to be flat. $S_n$ are the symmetric polynomials, given by
\begin{align*}
S_0(M)&=1\\
S_1(M)&=[M] \\
S_2(M)&={1\over 2!}\left( [M]^2-[M^2]\right)\\
S_3(M)&={1\over 3!}\left( [M]^3-3[M][M^2]+2[M^3]\right)\\
S_4(M)&= {1\over 4!}\left( [M]^4-6[M]^2[M^2]+8[M][M^3]+3[M^2]^2-6[M^4]\right) \nn \\
S_5(M)&= {1\over 5!}\left( [M]^5-10[M]^3[M^2]+15[M][M^2]^2+20[M]^2[M^3]-20[M^2][M^3]-30[M][M^4]+24[M^5]\right), \nn
\end{align*} 
where $[M]$ denotes the trace of $M^{\mu}{}_{\nu}$.
The term in \eqref{5ddRGT} proportional to $\beta_5$ does not contribute to the equations of motion. Furthermore, imposing that flat spacetime (or $M^4 \times S^1$) is a solution and that $m$ is the graviton mass gives the conditions
\begin{align}
\beta_0 + 4 \beta_1 +6 \beta_2 +4 \beta_3 + \beta_4 & =0 \label{beta1}, \\ 
\beta_1+3 \beta_2 + 3 \beta_3 + \beta_4 & = 8 \label{beta2},
\end{align}
which leaves a three-parameter family of theories. 

To perform the general dimensional reduction we will consider a simple case of \eqref{5ddRGT} given by the  ``minimal model" of \cite{Hassan:2011vm}, which corresponds to $\beta_1=8$, $\beta_0= -32$, and the other $\beta_n$ vanishing. We also find it convenient to use the vielbein formulation, following \cite{Hinterbichler:2012cn}.
The action for the five-dimensional minimal model in vielbein form is
\be \label{minmodel}
S = \frac{M_{(5)}^3}{2} \int d^4 x \int_0^L dy |E| \left( R(E) + 8m^2 -2m^2 E^{M}{}_A \delta_M{}^A\right),
\ee
where $E_{M}{}^A $ and $\delta_M{}^A$ are the vielbeins for $G_{MN}$ and $\eta_{MN}$ respectively,
\be
G_{MN} = E_{M}{}^A E_{N}{}^B \eta_{AB}, \quad \eta_{MN} = \delta_{M}{}^A \delta_{N}{}^B \eta_{AB},
\ee 
$|E|$ is the determinant of $E_{M}{}^A$, and $E^M{}_A$ is the inverse vielbein
\be
E^M{}_A E_N{}^A = \delta^M_N.
\ee
The vielbein and inverse vielbein corresponding to \eqref{KKansatz} are 
\be \label{vielbein}
 \hat{E}_M{}^{ A} = \left( \begin{array}{cc}
e_{\mu}{}^{a}(x) & \phi(x) A_{\mu}(x) \\ 
0 & \phi(x) \end{array} \right), \quad
\hat{E}^M{}_{ A} = \left( \begin{array}{cc}
e^{\mu}{}_{a}(x) & 0 \\ 
- e^{\nu}{}_a(x) A_{\nu}(x) & \phi^{-1}(x) \end{array} \right),
\ee
where $e_{\mu}{}^a$ and $e^{\mu}{}_a$ are the vielbein and inverse vielbein for $g_{\mu \nu}$.
Putting a general vielbein in the diagonal form \eqref{vielbein} requires a local rotation of four-dimensional spacetime into the fifth dimension. A general vielbein may thus be written as
\be \label{generalvielbein}
E_M{}^{ A} = \Lambda^{*A}{}_B \hat{E}_M{}^{ B},
\ee
where $\hat{E}_M{}^{ B}$ is upper triangular and $\Lambda^{*A}{}_B$ defines a local rotation. We can parameterize the required rotation in terms of a vector $B^a$ by writing the generator as
\be
\omega^A{}_B =  \left( \begin{array}{cc}
0 & B^a \\ 
-B_b & 0 \end{array} \right),
\ee
where $\Lambda^{*A}{}_B= \left( e^{\omega} \right)^{A}{}_B$ and $B_b \equiv \eta_{b a} B^a$.
Defining $B^2 \equiv B_{a} B^{a}$ and $|B| \equiv \sqrt{B^2}$, we can write the rotation matrix as
\be \label{rotation}
\Lambda^{*A}{}_{B} = \delta^A{}_B + \frac{(\omega^2)^A{}_B}{B^2} \left(1- \cos |B| \right) + \omega^A{}_B \frac{\sin |B|}{|B|}.
\ee

We can now find the dimensionally reduced action by plugging \eqref{generalvielbein} into \eqref{minmodel} and using \eqref{rotation}. The Einstein-Hilbert kinetic term is locally Lorentz invariant so $B_{a}$ only appears through the potential. Defining $B^{\mu} = e^{\mu}{}_a B^a$, so that $B^2=B_{\mu} B^{\mu}$, we can write the dimensionally reduced theory as
\begin{align}
S = & \frac{M_{(4)}^2}{2}  \int d^4 x |e| \phi  \left[ R(e) - \frac{1}{4} \phi^2 F_{\mu \nu}^2+8 m^2 \right. \nn\\
& \left. -2m^2 \left( \delta_{\mu}{}^a e^{\mu}{}_a + e^{\nu}{}_a \delta_{\mu} {}^a \frac{B^{\mu} B_{\nu} }{B^2}( \cos |B| -1) + \phi^{-1} \cos |B| +  A_{\mu} B^{\mu} \frac{\sin |B|}{|B|}\right) \right],\label{dimreddrgtac}
\end{align}
where $\delta_{\mu}{}^a$ is the vielbein for the four-dimensional flat fiducial metric.

The field $B_{\mu}$ appears in \eqref{dimreddrgtac} without derivatives and is thus an auxiliary field that should be integrated out using its equations of motion. We will not attempt to integrate out $B$ exactly, but it can be done to any desired order in powers of the fields. For example, to quadratic order we can write
\be
S = \int d^4 x \left[ \epsilon(h)_{m^2} - \frac{1}{4} F_{\mu \nu}^2 -2m^2 (A_{\mu} B^{\mu}-B^2) -(\partial \psi)^2 -m^2 \psi^2 \right],
\ee
where $\phi \equiv  e^{-\psi}$ and we have demixed $h_{\mu \nu}$ and $\phi$. Eliminating $B$ using its equations of motion gives
\be
S = \int d^4 x \left[ \epsilon(h)_{m^2} - \frac{1}{4} F_{\mu \nu}^2 -\frac{1}{2}m^2 A^2 -(\partial \psi)^2 - m^2 \psi^2 \right],
\ee
which describes a massive graviton, vector and scalar with mass $m$. This agrees with the zero-mode spectrum obtained in Section \ref{sec:N1} for $N=1$ and $d=4$.

The theory \eqref{dimreddrgtac} is a massive scalar-vector-tensor theory that derives from the five-dimensional minimal model.  We can obtain a massive scalar-tensor theory by setting  $A_{\mu}=0$, which is classically consistent since $A_{\mu}$ does not appear linearly after integrating out $B_{\mu}$. When $A_{\mu}=0$, the auxiliary field's equations of motion are solved by $B_{\mu}=0$, which corresponds to the simple metric ansatz
\begin{equation} \label{simpleansatz}
G_{A B}(x) = \left( 
\begin{array}{ccc}
g_{\mu \nu}(x) & 0 \\
0 & \phi^2(x)
\end{array} \right).
\end{equation}
Using the ansatz \eqref{simpleansatz}, we can readily find the associated dimensional reduction for the full three-parameter higher-dimensional dRGT theory \eqref{5ddRGT}. The result of this is
\be 
S = \frac{M_{(4)}^2}{2} \int d^4 x \sqrt{-g} \phi \left( R(g) - \frac{m^2}{4} \sum_{n=0}^4 \left( \beta_n + \frac{\beta_{n+1}}{\phi} \right) S_n \left( \sqrt{g^{-1} \eta} \right) \right),
\ee
where we have dropped the $S_5$ term since it becomes a total derivative in four dimensions. Again, the parameter $\beta_5$ does not contribute to the equations of motion. 
 In Einstein frame the action is
\be \label{scalartensor}
 S =  \frac{M_{(4)}^2}{2}  \int d^4 x \sqrt{-\tilde{g}}  \left( \tilde{R} -\frac{3}{2} (\partial \psi)^2 - \frac{m^2}{4} \sum_{n=0}^4 e^{\frac{(4-n)}{2}  \psi} \left( \beta_n e^{-\psi} + \beta_{n+1} \right) S_n \left( \sqrt{\tilde{g}^{-1} \eta} \right) \right),
\ee
where $\tilde{g}_{\mu \nu} \equiv \phi g_{\mu \nu}$ and $\phi \equiv e^{- \psi}$. The action  \eqref{scalartensor} defines a theory with three parameters in addition to the graviton mass and cosmological constant\footnote{In deriving \eqref{scalartensor} we used a flat fiducial metric and imposed \eqref{beta1} and \eqref{beta2}, but the final result makes sense for fiducial metrics and backgrounds that are not flat. }.  At quadratic order around flat spacetime, \eqref{scalartensor} describes a free graviton and scalar, both with mass $m$. Because \eqref{scalartensor} arises as a consistent truncation of dRGT ghost-free massive gravity, we expect it to also be ghost free.  It is an example of a  type of mass-varying massive gravity \cite{D'Amico:2011jj,Huang:2012pe} in which the graviton mass and dRGT parameters are promoted to functions of a scalar field. Cosmological perturbations of this general class of theories were studied in \cite{Gumrukcuoglu:2013nza}.

\section{Conclusions}

We have derived the linearized KK spectrum of a generic massive spin-2 field propagating on an Einstein direct product space.
We found that in the case where the higher-dimensional graviton is partially massless, its zero mode does not lead to a lower-dimensional partially massless graviton.  In this sense, dimensional reduction destroys partial masslessness.  The higher-dimensional partially massless symmetry instead acts to remove a tower of scalar fields from the spectrum.  Among the scalars which are removed is the zero mode, i.e. the radion.  The radion, like other moduli, is a light scalar and we see no sign of such light scalars in nature, so in realistic models they must be stabilized, or screened by nonlinearities \cite{Vainshtein:1972sx,Khoury:2003aq,Hinterbichler:2010es}.  One can speculate that partial masslessness might be useful as a mechanism for removing such moduli without resort to stabilization or screening.

In addition, we studied the zero-mode sector of the dimensional reduction of ghost-free dRGT massive gravity on a circle. This leads to a three-parameter family of massive scalar-tensor-vector theories, which we expect to be ghost free, and whose scalar-tensor sectors are examples of mass-varying massive gravity.  It would be interesting to further explore the consequences of these theories and to try to prove their ghost-freedom directly.



{\bf Acknowledgements:} J.B. would like to thank the Perimeter Institute for hospitality while this work was under way.  Research at Perimeter Institute is supported by the Government of Canada through Industry Canada and by the Province of Ontario through the Ministry of Economic Development and Innovation.

\appendix
\section{Diagonalization Transformations} 

In this appendix we give the transformations needed to diagonalize the action and discuss some special cases. When there are multiple demixing transformations they are performed in the order listed. We define
\[
m^2_k \equiv m^2+ \lambda_k - \frac{2R_{(d)}}{d},
\]
where $k= \mathcal{I},$ $0,$ $i$, or $a$ and $\lambda_0=0$.
 
\subsubsection*{Transformations for $\mathcal{L}_i$}
To diagonalize $\mathcal{L}_i$ for $i \neq \rm{Killing}$ we transform
\[
A_{\mu}^i \rightarrow A_{\mu}^i +\frac{1}{m_i^2} \left(m^2_i-m^2 \right)\nabla_{\mu} \phi^i.
\]
Since $\lambda_{i \neq \rm{Killing}} > 2 R_{(d)}/d$ for $i \neq \rm{Killing}$ we have $m_i>m$ and hence this transformation is well defined.

\subsubsection*{Transformations for $\mathcal{L}_0$}
When $m$ is not equal to $m_{(d)\rm{PM}}$ and $d>1$, $\mathcal{L}_0$ is diagonalized by 
\be \label{L0transf}
h_{\mu \nu}^{0} \rightarrow h_{\mu \nu}^0 - \frac{1}{(d-1)(m^2-m^2_{(d)\rm{PM}})}\left( \left( m^2 - \frac{R_{(d)}}{d} \right)g_{\mu \nu} \phi^0 - \nabla_{\mu} \nabla_{\nu} \phi^0 \right).
\ee
This transformation is not defined when $m=m_{(d)\rm{PM}}$. This is because the $h_{\mu\nu}^0$ kinetic term in the original action acquires a PM symmetry and hence the $h^0 \phi^0$ term cannot be unmixed by a transformation of the form \eqref{L0transf}, since this demixing transformation takes precisely the form of a PM gauge transformation. Instead, we can first eliminate the derivative scalar-tensor term by transforming
\be
{h}_{\mu \nu}^{0} \rightarrow h_{\mu \nu}^0 + \frac{1}{m^2} \nabla_{\mu} \nabla_{\nu} \phi^0,
\ee
then we can remove the scalar kinetic term using a PM transformation, and finally the ${h}^0 \phi^0$ term can be removed by redefining $\phi^0$. This leaves 
\be \label{PMzeromode}
\mathcal{L}_0 =  \epsilon({h}^0)_{m^2_{(d)\rm{PM}}} - \frac{1}{4} m^2_{(d)\rm{PM}} |{h}_0|^2 + \frac{m^2_{(d)\rm{PM}}}{(d-2)^2} |\phi^0|^2.
\ee
The scalar is now decoupled and its equations of motion just set $\phi^0 =0$, so we can drop it, giving \eqref{PMzeromode1}.

\subsubsection*{Transformations for $\mathcal{L}_{a}$}

 First we consider $N>1$. When $d>1$ and $m^2_a \neq 0$, we diagonalize $\mathcal{L}_a$ with the following transformations
\begin{align}
{h}_{\mu \nu}^a \rightarrow & {h}_{\mu \nu}^a - \frac{N-1}{N(d-1)(\lambda_a+m^2-m^2_{(d)\rm{PM}})}\left[ \lambda_a \left(\lambda_a - \frac{R_{(N)}}{N-1}\right) \left( g_{\mu \nu} {\phi}^a + \frac{d-2}{\lambda_a + m^2}\nabla_{\mu} \nabla_{\nu} {\phi}^a\right) \right. \nn\\
& \left. +\left(\lambda_a + \frac{Nm^2 - R_{(N)}}{N-1} \right) g_{\mu \nu} \bar{\phi}^a - \frac{m^2N +(D-2)\lambda_a}{(N-1)(\lambda_a+m^2)} \nabla_{\mu} \nabla_{\nu} \bar{\phi}^a \right] + \frac{\lambda^a}{\lambda_a+m^2} (\nabla_{\mu} A_{\nu} + \nabla_{\nu} A_{\mu})\,, \\
{A}_{\mu}^a \rightarrow & A_{\mu}^a +\frac{1}{N m_a^2} \left( (N-1)\left( \lambda_a - \frac{R_{(N)}}{N-1}\right) \partial_{\mu} {\phi}^a- \partial_{\mu} \bar{\phi}^a\right).                                                                                                                                                                                                                                                                                                                                                                                                                                                                                                                                                                                                                                                                                                                                                                                                                                                                                                                                                                                                                                                                                                                                                                                                                                                                                                                                                                                                                                                                                                                                                                                                                                                                                                                                                                                                                                                                                                                                                                                                                                            
\end{align}
When $a \neq \rm{conformal}$ we also transform 
\begin{equation} \label{phitrans}
{\phi}^a \rightarrow \phi^a- \frac{(D-2)m_a^2+m^2d}{ \lambda_a\left(\lambda_a(D-2)-2N \frac{R_{(d)}}{d}\right)+\left( 2 \lambda_a + m^2 N \right) (d-1) \left(m^2-m^2_{(d)\rm{PM}}\right)} \bar{\phi}^a.  
\end{equation}
These transformation give the diagonalized Lagrangian \eqref{diaglanoncon}. The coefficients $\sigma_a$ and $\bar{\sigma}_a$ for $a \neq \rm{conformal}$ are given by\footnote{The transformation \eqref{phitrans} is undefined for certain $\lambda_a$ when $m<m_{(N)\rm{PM}}$. At these points the determinant of the kinetic matrix \eqref{determinant} is negative and the Lagrangian can be put in the form \eqref{diaglanoncon} with $\sigma_a=1$, $\bar{\sigma}_a=-1$.
}
\be
\sigma_a =  \frac{(N-1) \lambda_a\left(\lambda_a - \frac{R_{(N)}}{N-1} \right)\left(\lambda_a\left(\lambda_a(D-2)-2N \frac{R_{(d)}}{d}\right)+ \left( 2 \lambda_a + m^2 N \right) (d-1) \left(m^2-m^2_{(d)\rm{PM}}\right)\right)}{2N^2(d-1)m_a^2 \left(\lambda_a+m^2-m^2_{(d) \rm{PM}}\right)}\, ,
\ee
and
\be \label{determinant}
\sigma_a \bar{\sigma}_a = \frac{\lambda_a m^2(N-1)(D-1)\left(\lambda- \frac{R_{(N)}}{N-1}\right)(m^2-m^2_{(D)\rm{PM}})}{4N^2(d-1)m_a^2(m^2+\lambda_a-m^2_{(d)\rm{PM}})}.
\ee
When $a = \rm{conformal}$, $\sigma_a =0$ and
\be
\bar{\sigma}_a = \frac{(m^2-m^2_{(D)\rm{PM}})}{ m^2_a (m^2+\frac{R_{(D)}}{D}\frac{(D-2)}{(d-1)(N-1)})}\frac{m^2(D-1)}{2N(d-1)}.
\ee
When $m_a=0$ the Lagrangian is diagonalized by transforming $h_{\mu \nu}$ as above and then transforming
\begin{align}
\bar{\phi}^a \rightarrow & \bar{\phi}^a- (N-1)(m^2-m^2_{(N) \rm{PM}} )\phi^a, \\
A_{\mu}^a \rightarrow  &A_{\mu}^a - \frac{(N-1)\left( m^2-m^2_{(N) \rm{PM}} \right)\left( m^2  + (D-2) \frac{R_{(d)}}{d} \right)}{m^2 d \left( m^2-\frac{ 2 R_{(d)}}{d} \right)} \partial_{\mu} \phi^a \nn\\ &-\frac{1}{4} \left( \frac{(N-2)(D-2)}{d m^2 N}-\frac{D}{dm^2-2 R_{(d)}}+\frac{d-2}{NR_{(d)}}  \right) \partial_{\mu} \bar{\phi}^a, \\
\bar{\phi}^a \rightarrow & \bar{\phi}^a  - N \nabla^{\mu} A_{\mu}^a.
\end{align}
Note that when $a =\rm{conformal}$, $m_a=0$ corresponds to $m=m_{(N) \rm{PM}}$. These transformations leave \eqref{conformspecial2} and a nondynamical term proportional to $|\bar{\phi}^a|^2$.

When $d=1$, $\mathcal{L}_{a=\rm{conformal}}$ is nondynamical. $\mathcal{L}_{a \neq \rm{conformal}}$ can be simplified by transforming $h^a \rightarrow h^a + 2\nabla_{\mu}A^{\mu}_a$ and $\bar{\phi}^a \rightarrow \bar{\phi}^a-\lambda^a \phi^a$. Both $h_{\mu \nu}^a$ and $A_{\mu}^a$ then appear algebraically and can be integrated out using their equations of motion, which leaves \eqref{noncond1}.

Lastly, consider $N=1$, i.e. $\mathcal{N}$ is the circle with radius $r$. In this case $R_{(N)}=0$, all scalars are conformal scalars, and the scalar Laplacian eigenvalues are given by $\lambda_a= a/r^2$, where $a$ is a positive integer. Assuming $d>1$, to demix we must transform $h^a_{\mu \nu}$ as above and transform
\begin{align*}
{A}_{\mu}^a \rightarrow & A_{\mu}^a - \frac{1}{\lambda_a+m^2} \partial_{\mu} \bar{\phi}^a,
 \end{align*} 
 which gives the demixed Lagrangian \eqref{conformalN1}.

\bibliographystyle{utphys}
\addcontentsline{toc}{section}{References}
\bibliography{PM-KK-arxiv-v2}

\end{document}